# Permeation behaviour of hydrogen isotopes in molten FLiBe (2LiF-BeF$_2$): Identifying sources of uncertainty and associated measurement challenges

Abhishek Saraswat [a], Weiyue Zhou [a*], Nayoung Kim [a], Jaron F. Cota [a], Guiqiu Zheng [b], Alexander A. Khechfe [b], Caroline S. Barthel [b], Michael P. Short [a], Rémi Delaporte-Mathurin [c], Kevin B. Woller [c*]

[a] Department of Nuclear Science and Engineering, Massachusetts Institute of Technology, Cambridge, MA 02139, USA
[b] Commonwealth Fusion Systems, Devens, MA 01434, USA
[c] Plasma Science and Fusion Center, Massachusetts Institute of Technology, Cambridge, MA 02139, USA

*Corresponding authors: wyzhou@mit.edu, kbwoller@mit.edu

**ABSTRACT**

This paper presents results from systematic investigations conducted in the HYPERION facility to quantify permeabilities of hydrogen isotopes in FLiBe over a temperature range of 773K - 973K. To address the knowledge gap resulting from widely scattered transport parameters reported in the earlier studies, HYPERION experiments incorporate specific provisions to probe unsubstantiated assumptions, including a one-dimensional permeation, membrane surface coverage by the salt, and ideal wettability of the salt-metal interface. Suppression of the isotopic transport behavior for metal-side charging highlights the permeation barrier characteristics of a bubble-laden Ni-FLiBe interface, impacting permeability by up to 77%. This work attempts to resolve this issue via salt-side charging and informs constraints on the permeant charging methodology for future studies. The initial insights from HYPERION experiments call into question the design choices and estimated transport properties in earlier works. This study provides a plausible explanation for the observed scatter in H/D/T transport in FLiBe.

## 1. INTRODUCTION

An effective Fusion Power Plant (FPP) design mandates closing the fusion fuel cycle, characterized by a Tritium Breeding Ratio (TBR) greater than unity. A well-devised arrangement of functional materials could help achieve this aim through neutron multiplication and tritium breeding. In particular, Li-containing liquid metals (eutectic PbLi, Li) and molten salts (FLiBe, FLiNaBe, etc.) are the preferred liquid breeder candidates for these applications. Additional technological merits such as high boiling points, resistance to radiation damage, better heat extraction capabilities, and the possibility of tritium extraction outside the blanket make liquid breeders preferable to solid breeders [1-2]. However, high electrical conductivities of liquid metals make them inherently prone to exorbitantly large Magneto-Hydrodynamic pressure drops ($\Delta P_{MHD}$), exerting substantial demands on the required pumping power and associated structural integrity [2-3]. The validation of designs employing electrically insulating coatings and flow channel inserts to alleviate $\Delta P_{MHD}$ remains confined to lab-scale prototypical tests [4-5].

Molten salts, in contrast, exhibit orders of magnitude lower electrical conductivities and, therefore, alleviate concerns related to $\Delta P_{MHD}$ [6]. FLiBe, a fluoride salt with a 2:1 molar composition of LiF and BeF$_2$, is gaining immense focus in the fusion community owing to its specific advantages over candidate liquid metal breeders and ceramic breeders. A higher achievable TBR with moderate $^6$Li enrichment and its neutron moderation characteristics make FLiBe an attractive choice over other breeder salts [7-8].

FLiBe-based Liquid Immersion Blanket (LIB) concepts adopted by ARC-class FPPs aim to drive a global TBR > 1 through minimization of structural materials within the bulk blanket assembly [9]. Further, a robust accounting of tritium yield with a low uncertainty is one of the key metrics to evaluate the blanket performance towards self-sufficiency goals [10]. Realization of these efforts necessitates a detailed understanding of tritium migration within the blanket and associated critical components. An experimental quantification of transport parameters with improved accuracy is also a requisite for the validation of transport models representing complex integrated systems in an FPP [11]. Moreover, the design of a Tritium Extraction System (TES) - one of the critical components addressing fuel recycling and tritium inventory management – is governed by an exhaustive characterization of tritium transport behaviour within FLiBe over the temperature regime representative of fusion breeding blankets.

Despite the importance of permeation data, experimental studies for tritium permeation in FLiBe remain scarce owing to Be toxicity, limited tritium availability, and associated material handling and regulatory requirements [12-15]. A handful of available studies utilizing H$_2$ and D$_2$ as surrogate isotopes also depict a wide scatter in the estimated transport parameters. For instance, the solubility of H$_2$ in FLiBe as estimated by Malinauskas *et al.* (1974) is two orders of magnitude lower than that estimated by Nakamura *et al.* (2015), although both studies report H$_2$ transport in its molecular form [16-17]. Similarly, the estimated H$_2$ permeability in FLiBe differs by a factor of ~ 4 in sequential studies conducted by the same group (Nakamura *et al.*, 2015 and Nishiumi *et al.*, 2016) using similar apparatus [17-18]. In contrast, the permeability of tritium measured by Calderoni *et al.* (2008) is an order of magnitude higher than the $H_2$ permeability measured by Nakamura *et al.* (2015) [17,19]. It is also worth highlighting that the tritium diffusivity estimates made by Calderoni *et al.* (2008) remain higher than all the corresponding estimates made for $H_2$ and $D_2$ in other studies with FLiBe. Such inconsistent behaviour in the transport properties of $H_2$ and its isotopes could be attributed to a variety of factors, including salt chemistry, corrosion, and a simultaneous presence of multiple hydrogen isotopes [20-21], to name a few. Finally, much of the existing literature lacks rigorous quantification of experimental uncertainties in the estimated transport parameters.

The present study is conducted in an upgraded version of the HYPERION (HYdrogen PERmeatION) facility [22] to achieve the following pre-defined objectives: (i) to quantify permeation behaviour of hydrogen isotopes in FLiBe, (ii) to generate an independent data-set for possible comparisons against available results from earlier studies, and (iii) to shed some light on the possible reasons behind a significant scatter in the transport properties observed in earlier studies. In this study, the FLiBe sample is melted inside a chemically compatible Ni-200 permeation cell, externally coated with ~250 µm thick thermally-sprayed $Al_2O_3$, to generate a salt layer for permeation experiments. The permeation cell placed inside a regularly purged glovebox (GB) provides an inert environment to reduce uncertainties related to the metallic/non-metallic impurities, while simultaneously addressing possible errors associated with the background concentration build-up. Intermittent visual inspections of the salt layer over the metallic membrane eliminate ambiguities related to incomplete surface coverage. A proper accounting of experimental uncertainties and an explicit mention of the adopted units for the parameters provide a necessary reference for comparisons in future works.

## 2. MATERIALS AND METHODS

### 2.1. System design description

A process flow diagram for the HYPERION facility and an image of the assembled permeation cell within GB are shown in **Fig.1(a,b)**, respectively. The permeation cell consists of a 78 mm inner diameter cylindrical chamber, made of Ni, separated into upstream and downstream volumes ($V_{up}$ and $V_{down}$) by a Ni membrane. To impart the necessary structural integrity for high-temperature operations, the vessel and membrane are machined in place from a solid Ni-200 bar to achieve a seamless construction. The membrane and vessel walls are 2 mm thick with a mill tolerance of ±12.5% and a surface roughness of 3.2 µm. The only welded components in the permeation cell assembly are the top and bottom cover lids with gas inlet/outlet ports. The port tubes on the top lid additionally provide access for FLiBe inventory feed and membrane inspections using an inspection camera. The vessel assembly is placed inside a negative-pressure GB with an inert (Ar) environment and active control of $O_2$ and $H_2O$ levels below 1 ppm each. Ultra-High Purity (UHP) grade gaseous Ar and $Q_2$ ($Q \equiv H/D$) are used in the downstream and upstream circuits, respectively, with inline high-performance traps (Make: RESTEK) enabling $O_2$/water-vapour removal to low ppb levels before entering the permeation cell.

The permeated $Q_2$ in $V_{down}$ is swept through by an Ar stream at a constant flow rate and is measured using a calibrated Gas Chromatograph (GC, Agilent 8890) equipped with a Thermal Conductivity Detector (TCD). A nearly tenfold difference in the thermal conductivities of Ar and $Q_2$ allows for a practically achievable detection limit as low as 20 ppm, below which distortions in the concentration peaks measured by the TCD prohibit a meaningful interpretation. The vessel overpressure protection is addressed through proportional Relief Valves (RVs) located at the inlet lines, while potential Be migration towards the GC is arrested through 2-µm particulate filters (FF) installed at the outlet lines for both circuits. A Digital Mass Flow Controller (DMFC, Model: Omega FMA5504A) in conjunction with a Non-Return Valve (NRV) helps maintain an undisturbed sample flow towards the GC. The upstream $Q_2$ concentration is maintained constant using an assembly of a back pressure regulator and digital pressure gauges. Before experiments, K-type thermocouples are used for membrane surface temperature calibrations for dry runs (without salt) and FLiBe runs, while the assembly heating is achieved through a vertical split-tube furnace (~ 1.8 kW). Since the $Q_2$ Permeation Reduction Factor (PRF) for the $Al_2O_3$ coating applied on the permeation cell is not known *a priori*, the permeability estimations for Ni and FLiBe are reported within the bounds of unity and infinite PRFs. PRF is defined as the ratio of permeation flux through an uncoated (bare) substrate to the permeation flux through a coated substrate. Thus, a unity PRF effectively refers to a coating with no permeation barrier effect, and an infinite PRF refers to an ideal coating with no permeation across it. The intrinsic hydrogen permeabilities are expected to lie within these limits.

A portable electrochemical $H_2$ sensor (Model: Industrial Scientific, Pro GasBadge) installed inside the GB provides information on $Q_2$ concentration build-up within the GB volume ($V_{GB}$) over time. A regular inert gas (Ar) purging of the $V_{GB}$ helps maintain a $Q_2$ partial pressure in the GB below the $Q_2$ partial pressure in the $V_{down}$ to arrest any reverse permeation towards $V_{down}$. Accounting and correction for the radial permeation losses are discussed in detail under *Section-2.2*. Calibrations for the TCD and $H_2$ monitor using $H_2$-Ar / $D_2$-Ar calibration gas mixtures enable quantitative comparisons of the transport behaviours for the two isotopes. For calibration purposes, $H_2$-Ar (balance) mixtures used are 49.5 ppm ± 5%, 486 ppm ± 2% and 1992 ppm ± 2%, while $D_2$-Ar (balance) mixtures used are 52.4 ppm ± 2% and 499 ppm ± 2%.

The FLiBe salt used in this study is sourced from SaltGen Inc. (USA). The vendor-provided test certificates list an oxide concentration below 50 ppm, although this is not independently verified. Measurements of cationic impurities in the salt, performed using Inductively Coupled Plasma Mass Spectrometry (ICP-MS), are presented in **Table-1**. For the ICP-MS analysis, approximately 100 mg of FLiBe is dissolved in a mixed acid solution prepared by combining 7 mL of 16 M $HNO_3$ with 5 mL of 5 vol.% $HNO_3$, the latter containing 10 ppm Tb as an internal standard (Inorganic Ventures). After a static digestion for 24 h duration, the samples are further digested using an Ultrawave Microwave Digestion system (Milestone) with a 15 min ramp to 230°C followed by 40 min hold at 230°C. The digested solutions are then diluted to obtain 5 vol.% HNO$_3$ samples. Three replicate samples are prepared, and each is analyzed in triplicate using Agilent 7900 ICP-MS. Utilized instrumental settings are as follows: RF power (1550 W), RF current (1.80 V), nebulizer gas flow rate (0.95 L/min), and dilution gas flow rate (0.11 L/min). Helium gas collision mode is employed to avoid polyatomic interferences. Calibration for the ICP-MS system is performed using multi-element standard solutions (Inorganic Ventures) at concentrations of 10, 20, 50, 100, 200, 500, 1000, 2000, and 5000 ppb. Calibration curves with a correlation value greater than 0.995 are selected. The isotopes chosen for the analysis are $^{24}$Mg, $^{27}$Al, $^{44}$Ca, $^{52}$Cr, $^{55}$Mn, $^{56}$Fe, $^{59}$Co, $^{60}$Ni, $^{63}$Cu, $^{66}$Zn, $^{137}$Ba, and $^{208}$Pb. More details on the principle and procedures for ICP-MS measurements are available in our previous publication [23]. As mentioned under **Table-1**, the impurity levels with respect to transition metals, which are known to significantly alter the redox chemistry of the salt, either meet or exceed the target limits established by Molten Salt Reactor Experiment (MSRE) and contemporary studies [24-25]. The higher uncertainty values for some of the impurities in **Table-1** could be attributed to the technical difficulties encountered in a representative sampling of the salt [23]. Images of the as-sourced salt and cast cylindrical pellets are available in **Appendix-A: Supplementary data (Fig.S1)**.

Table-1: ICP-MS based impurity analysis for the FLiBe sample used in HYPERION studies

| Impurity | Mg | Al | Ca | Cr | Mn | Fe | Co | Ni | Cu | Zn | Ba | Pb |
|---|---|---|---|---|---|---|---|---|---|---|---|---|
| ppm (wt.) | 806.4 ± 11.3 | 143.5 ± 44.9 | < 7.2 | 2.1 ± 0.7 | 1.2 ± 0.1 | 21.8 ± 3.1 | < 0.1 | 15.7 ± 0.5 | < 0.1 | 82.8 ± 35.9 | 2.6 ± 0.3 | 0.3 ± 0.1 |

## 2.2. Operational procedure

The operating ranges for different process variables are listed in **Table-2**. Sequential Ar purging of both $V_{up}$ and $V_{down}$ is followed by vessel heating and stabilization at the required investigation temperature. The $V_{up}$ is then purged with a ~ 20 sccm flow of 100% Q$_2$ at the required investigation pressure, gradually displacing the existing Ar through dilution purging [26]. This is followed by maintaining a minimal Q$_2$ flow at ~ 5 sccm in $V_{up}$.

**Fig.2** depicts possible permeation paths for the as-built HYPERION system. Permeation along path-I (axial) is desired, but a complete seamless construction may inevitably allow radial permeation along path-II and path-IV if the coating has a finite PRF at operating conditions. The radial loss along path-IV is addressed using active Q$_2$ concentration measurements in $V_{GB}$ to correct for an ideal 1D permeation flux ($J_{axial}$). A low Q$_2$ partial pressure maintained in $V_{GB}$ arrests any back permeation towards $V_{down}$ (along path-III). The impact of bubbles at the Ni-FLiBe interface is explored by charging Q$_2$ at the salt side, effectively swapping $V_{up}$ and $V_{down}$ in the configuration presented in **Fig.2**.

Table-2: Operating ranges for process variables

| Parameter | Units | Operating range |
|---|---|---|
| Temperature ($T$) | K | 773 - 973 |
| Upstream pressure ($P_{up,Q2}$) | MPa(a) | 0.1 - 0.13 |
| Downstream pressure ($P_{down}$) | MPa(a) | 0.1 - 0.12 |
| Upstream flow ($F_{up, Q2}$) | sccm | ~ 5, ~ 20 |
| Downstream flow ($F_{down, Ar}$) | sccm | 29.15 |

If $Ø_{Ni}$ and $Ø_{FLiBe}$ represent the respective permeabilities for the Ni membrane and the FLiBe salt layer, $L_{Ni}$ and $L_{FLiBe}$ are their respective thicknesses, $P_{up,Q2}$, $P_{down,Q2}$ and $P_{GB,Q2}$ are the partial pressures of H$_2$/D$_2$ in $V_{up}$, $V_{down}$ and $V_{GB}$, respectively, while $P_{interface,Q2}$ denotes the H$_2$/D$_2$ pressure at the Ni-salt interface, then the effective permeabilities for the bounding PRF cases can be estimated as described below:

**Case-1: Q$_2$ charging at the metal side:**

*a) PRF for the coating is unity:*

In this extreme case, $J_{radial} \neq 0$ along path-IV.

For both dry run (no salt layer) and FLiBe run steady state cases:

$$J_{axial} = J_{measured} + J_{radial} \quad (1)$$

$$J_{radial} = \frac{Ø_{Ni}}{L_{Ni}} \left( \sqrt{P_{down,Q2}} - \sqrt{P_{GB,Q2}} \right) \quad (2)$$

For investigations without salt ($L_{FLiBe} = 0$):

$$J_{axial} = \frac{Ø_{Ni}}{L_{Ni}} \left( \sqrt{P_{up,Q2}} - \sqrt{P_{down,Q2}} \right) \quad (3)$$

Arranging eq. (1), (2) and (3):

$$Ø_{Ni} = \frac{J_{measured} \times L_{Ni}}{(\sqrt{P_{up,Q2}} - 2\sqrt{P_{down,Q2}} + \sqrt{P_{GB,Q2}})} \quad (4)$$

For investigations with FLiBe:

$$J_{axial} = \frac{Ø_{FLiBe}}{L_{FLiBe}} \left( P_{interface,Q2} - P_{down,Q2} \right) \quad (5)$$

Arranging eq. (1), (2) and (5):

$$Ø_{FLiBe} = \frac{(J_{measured} + J_{radial}) \times L_{FLiBe}}{(P_{interface,Q2} - P_{down,Q2})} \quad (6)$$

where $J_{radial}$ at a given temperature is calculated using eq. (2) with $Ø_{Ni}$ estimated from the dry runs.

$P_{interface,Q2}$ is given by

$$P_{interface,Q2} = (\sqrt{P}_{up,Q2} - \frac{J_{axial} \times L_{Ni}}{Ø_{Ni}})^2 \quad (7)$$

where the second component on the right-hand side of eq. (7) represents the $Q_2$ pressure drop across the Ni membrane.

*b) PRF for the coating is infinite:*

In this extreme case, the coating is ideal and, therefore, $J_{radial} = 0$, resulting in $J_{axial} = J_{measured}$. Therefore, $Ø_{Ni}$ and $Ø_{FLiBe}$ can be estimated from eq. (3), (5) and (7) by substituting $J_{measured}$ for $J_{axial}$.

$$Ø_{Ni} = \frac{J_{measured} \times L_{Ni}}{(\sqrt{P}_{up,Q2} - \sqrt{P}_{down,Q2})} \quad (8)$$

$$Ø_{FLiBe} = \frac{J_{measured} \times L_{FLiBe}}{(P_{interface,Q2} - P_{down,Q2})} \quad (9)$$

where $P_{interface,Q2} = (\sqrt{P}_{up,Q2} - \frac{J_{measured} \times L_{Ni}}{Ø_{Ni}})^2 \quad (10)$

**Case-2: $Q_2$ charging at the salt side:**

Following similar derivations, permeabilities for this configuration can be expressed as:

*a) PRF for the coating is unity:*

$$Ø_{FLiBe} = \frac{(J_{measured} + J_{radial}) \times L_{FLiBe}}{(P_{up,Q2} - P_{interface,Q2})} \quad (11)$$

where $P_{interface,Q2} = (\sqrt{P}_{down,Q2} + \frac{J_{axial} \times L_{Ni}}{Ø_{Ni}})^2 \quad (12)$

*b) PRF for the coating is infinite:*

For this case, the expressions are again identified from eq. (11) and (12) using $J_{radial} = 0$, resulting in $J_{axial} = J_{measured}$.

Unlike [18], all the estimations in HYPERION investigations take into account a finite permeation resistance exhibited by Ni-200, resulting in $P_{interface,Q2} < P_{up,Q2}$ for metal-side charging, and $P_{interface,Q2} > P_{down,Q2}$ for salt-side charging. In some of the earlier studies, ambiguities in the adopted subscripts for the permeation flux are worth noting. For instance, Nakamura *et al.* (2015) uses $j_H$ and $j_{H2}$ interchangeably to depict steady state permeation flux in FLiBe, without an explicit definition of the species involved (mol-H or mol-$H_2$). Such inconsistencies hinder a proper comparison with the earlier efforts. In the present study, an emphasis is made to eliminate such ambiguities by adopting the H-atom permeation flux for Ni (a material obeying Sievert's Law) and $H_2$ permeation flux for FLiBe (a material obeying Henry's Law) [27-30]. This means the hydrogen/deuterium solubility follows $C \propto \sqrt{P_{up,Q2}}$ in Ni and $C \propto P_{up,Q2}$ in FLiBe, respectively, where $C$ represents the equilibrium concentration dissolved in the material. Results from both Case-1 and Case-2 are compared, within experimental uncertainty limits, to highlight the impact of an interface-limited transport regime resulting from bubble nucleation during Case-1 FLiBe runs.

## 3. RESULTS AND DISCUSSION

### 3.1. $H_2$ permeability estimations for Nickel

The vessel integrity is ascertained through regular pressure decay tests at room temperature, with $H_2$ and Ar introduced into $V_{up}$ and $V_{down}$, respectively. However, the $H_2$ concentration buildup within $V_{GB}$ is found to be non-negligible during high-temperature permeation tests, suggesting a finite permeation across the coating over the investigated temperature range. At each temperature, the permeation flux ($J_{measured}$) is measured for $P_{up,H2}$ of 0.11 MPa(a) and 0.13 MPa(a). One set at 0.11 MPa(a) is shown in **Fig.3**. For comparisons with experiments, simulations are performed in FESTIM (Finite Element Simulation of Tritium In Materials) – an open-source python code for multi-dimensional hydrogen transport modelling [31]. **Fig.3** also includes ideal 1D numerical results with the same $P_{up,H2}$, but a zero Dirichlet boundary condition applied downstream ($V_{down}$). The diffusivity and solubility parameters for these simulations are taken from Lee *et al.* (2013) [32].

The deviation between the experimental measurements ($J_{measured}$) and ideal 1D numerical results from FESTIM simulations increases with an increase in the operating temperature. This trend suggests a finite radial loss of permeation flux (path-IV) across the vessel walls at higher temperature with an increase in the Ni permeability. The longer time duration required to reach steady state in the experiments is primarily governed by a gradual displacement of Ar by $H_2$ in the $V_{up}$ and associated gas transfer lines. As the primary objective of this study is the estimation of permeabilities, a detailed analysis of the transient flux behaviour is outside the scope of the present investigations.

The estimated permeability values ($Ø_{Ni}$), accounting for the lost flux component as per eq.(4), are compared against the literature values [32-39] in **Fig.4**. Each data point from the present study is an average of two runs at different $P_{up,H2}$. A deviation of less than 3% is observed in the estimated permeabilities at any given temperature. The combined percentage uncertainty, derived from the root-sum-square of individual uncertainties, remains within ±13%. The Arrhenius plots from some of the studies are extrapolated in **Fig.4** to provide a comparison over the complete investigated temperature range. As expected, the reported permeabilities of deuterium in Ni [37-39], as shown in **Fig.4**, fall below the estimated permeabilities of hydrogen in Ni from HYPERION experiments over the complete temperature range of 773 K-973 K. A good agreement with previous studies corroborates the robustness of the methodology described under *Section-2.2*.

For the enveloping coating PRF cases, estimated hydrogen permeabilities for Ni-200 (units: H/m.s.Pa$^{0.5}$) are correlated to the temperature as per the following Arrhenius equations:

$$\emptyset_{H-Ni(PRF=1)} = 8.25 \times 10^{16} \exp(-46.7 \times 10^3/R_g T)$$

$$\emptyset_{H-Ni(PRF=\infty)} = 6.39 \times 10^{16} \exp(-45.3 \times 10^3/R_g T)$$

### 3.2. Permeation experiments for FLiBe: $H_2$ charging at the metal side

With the experimental methodology validated through dry runs, the preparation for the salt runs included melting ~ 37.5 g of FLiBe (in the form of cylindrical pellets) to generate a ~ 4 mm thick salt layer over the Ni membrane of the permeation cell (density reference taken from [40]). As shown in **Fig.5(a)**, visual inspection of the membrane revealed only a partial cross-sectional coverage. This effect could be attributed to a modest surface tension of FLiBe (~ 180 mN/m at 973 K) [40]. During some of the earlier attempts to create a ~ 2 mm thick FLiBe layer in another vessel of identical dimensions, the permeated $H_2$ concentration decreased by a mere ~ 40% as compared to the dry runs. This observation could be attributed to a preferential permeation path through the uncovered portion of the Ni membrane surface. Such an occurrence, if not corrected for, could be a potential source of error in the axial configuration set-ups, leading to a measured (apparent) salt permeability being orders of magnitude higher than the intrinsic permeability. Addition of more salt (total mass = 49.9 g), coupled with manual stirring, allowed complete membrane coverage with a ~ 5 mm thick FLiBe layer (**Fig. 5(b)**). Salt loading in powder form instead of pellet form could possibly help achieve a thinner salt layer. However, post-melting visual inspections to ensure complete surface coverage remain indispensable to mitigate possible uncertainties.

One set of measured flux for $P_{up,H2}$ = 0.10 MPa(a) and $P_{down}$ = 0.11 MPa(a) is shown in **Fig.6** for the metal-side $H_2$ charging configuration. The steady state flux values follow an expected trend of increase with an increase in temperature. It is worth noting that the permeation flux at a given temperature is two orders of magnitude lower than the measured permeation flux in dry runs (**Fig.3**).

However, a few runs displayed intermittent oscillatory behaviour in the magnitude of $J_{measured}$. These oscillations, lasting for a short duration, suggested an abrupt increase in the downstream $H_2$ concentration, followed by a return to the pre-oscillatory steady state. Detailed experimental investigations ascertained the appearance of these recurring oscillations when either the $H_2$ is charged at a higher pressure or the system is held at steady state for longer durations. One such instance illustrated in **Fig.7** suggests that the plateau observed before the oscillations is only an apparent steady state. Moreover, application of a higher Ar sweep pressure ($P_{down}$) is observed to suppress these oscillations. This behaviour is distinctly reminiscent of the behaviour of bubbles rising in a liquid column.

From these observations, it could be inferred that $H_2$ charging at the high permeability Ni side allows a larger permeation flux of H-atoms to the Ni-FLiBe interface, where the H-atoms combine to form $H_2$. Thereafter, a much lower permeability in the FLiBe layer prohibits $H_2$ transport at the same rate, resulting in the accumulation of $H_2$ at the interface when a non-ideal solid-liquid interface is present. A finite surface roughness may contribute to a non-ideal solid-liquid interface, which can allow a higher $H_2$ chemical potential to develop, facilitated by the supersaturation of the salt layer due to its much lower diffusivity than that of Ni. However, the nature of the bubbles in the present study remains unclear - specifically, whether the bubbles are formed of pure $H_2$ or of Ar + $H_2$. The bubbles of Ar + $H_2$ could be formed due to the entrainment of pre-existing Ar bubbles at the interface during salt melting under non-ideal solid-liquid interfacial conditions. In either case, the assumption of an ideal series-coupled Ni-FLiBe interface may not be valid, i.e. $J_{Ni} \neq J_{FLiBe}$. Further, the subsequent bubble growth could occur either through physical coalescence or through diffusion-driven mass transfer from smaller bubbles to the larger ones, reducing the surface energy through Ostwald ripening [41-42]. Finally, a larger bubble may get released under the influence of buoyancy, traversing across the salt layer and leading to a sudden increase in the measured downstream $H_2$ concentration, as observed in the GC signal. If the increased $H_2$ concentration is taken into account, this could lead to an overestimation of the salt permeability, as the bubbles traverse the salt layer under the effect of buoyancy (not diffusivity). It is, therefore, not representative of the transport phenomena relevant to permeability estimations. In contrast, neglecting the concentration from these oscillatory peaks would lead to an underestimation resulting from a reduced effective Ni membrane surface area available for permeation, as the interface sites covered with bubbles restrict the overall permeation by trapping H-atoms that have permeated through the Ni membrane.

In the present set-up, a visual inspection of the bubbles at high temperature is challenging, and a correction for effective permeation area calls for a precise quantification of the size of the bubbles residing at the interface. Fradera and Cuesta-López (2014) numerically studied the permeation barrier effect exhibited by nucleation of He bubbles at the PbLi/SS interface [43]. However, to the best of the authors' knowledge, no such numerical study exists for molten salt environments. Additionally, none of the previous experimental studies with FLiBe or FLiNaK reported the presence of bubbles, let alone their impacts, when $Q_2$ charging is performed at the metal side.

### 3.3. Establishing the impact of interface bubbles as permeation barriers: Isotopic masking effect for metal-side charging

To further establish the permeation barrier characteristics of a bubble-laden interface, separate runs are performed with $H_2$ and $D_2$ at 873 K with identical operating parameters. A comparison of $J_{measured}$ for the two isotopes is shown in **Fig.8**. The peaks near region 1 for the $D_2$ flux result from spontaneously detaching bubbles traversing through FLiBe, whereas the peaks near regions 2 and 3 result from forced bubble detachments through manual interventions (gentle vessel shaking). The critical observations from the $D_2$ permeation flux include the appearance of large peaks just after vessel shaking, and gradually diminishing peak magnitudes over the three instances of shaking. The following conclusions can be drawn from

**Fig.8**: (a) appearance of the peaks could not be attributed to the shaking process alone; (b) accumulation of gaseous $Q_2$ at the Ni-FLiBe interface occurs gradually resulting in bubble nucleation and growth; (c) gentle shaking of the vessel seems to release bubbles of certain characteristics (the last vessel shaking instance resulted in no peaks).

As the peak area generated by a TCD is primarily governed by the concentration and thermal conductivity of the eluted substance [44], a higher response factor is expected for $H_2$ compared to $D_2$. In contrast, the measured $D_2$ flux remains high over the complete test duration. Post-oscillations, the $D_2$ flux gradually reverts to an apparent steady state, still retaining a magnitude higher than the $H_2$ flux (encircled regions in **Fig.8**). This observation highlights a significant isotope masking effect on the permeation behaviour and suggests a continuous nucleation process at the Ni-FLiBe interface leading to an interface-limited transport regime. The initial observations from this study corroborate the claim of interfacial permeation impediment resulting from the formation of bubbles. However, detailed and rigorous investigations with possible interface imaging will be a focus of future studies.

### 3.4. Possible reasons for missing observations on bubble effects in earlier studies

A multitude of factors could drive the overall system behaviour towards an *apparent* suppression of bubble-induced permeation flux variations. In this context, some of the technical possibilities for earlier studies are discussed in detail here.

*3.4.1   Impact of impurities present in FLiBe*

The salt-metal interface coupling could be improved with a reduction in the surface tension of the salt itself. Sankar and Singh (2022) established that any level of oxide impurity drives the corrosion of structural materials in fluoride salts, while Gelbard *et al.* (2023) highlighted the role of such impurities as surfactants, lowering the surface tension of FLiBe [45-46]. Moreover, it is well established that the salt chemistry significantly influences H/D/T migration dynamics, while the corrosion of structural material(s) over long durations provides additional means of metallic impurity inclusions and material thinning, affecting the overall permeation [20].

In this context, the processed FLiBe used by Anderl *et al.* (2004) exhibited significant non-metallic impurities (O: 600 ppm and C: 45 ppm) [47], whereas no impurity analysis is available in the studies by Nakamura *et al.* (2015) and Nishiumi *et al.* (2016) - both of which utilized a radial permeation configuration with tertiary cylindrical Monel-400 tubes. Nakamura *et al.*, however, propose the presence of impurities and a non-uniform tube temperature distribution as possible sources of uncertainties. A reduced surface tension of FLiBe due to the impurities may result in good interface wetting, thereby suppressing possible bubble nucleation sites. However, in such cases, the observed transport behaviour may not be representative of the intrinsic permeability for pure FLiBe. In this work, the FLiBe layer imperfectly wetted the membrane and could have resulted in the formation of nano/micro-scale gas pockets at the solid-liquid interface, acting as $H_2$ trapping sites. This possibility could be correlated with the observations made in **Fig.5(a)**.

*3.4.2   Impact of system design choices on the observed permeation behaviour*

For specific cases, such as in Calderoni *et al.* (2008), where the FLiBe purity is similar to that of the HYPERION facility, design choices play a crucial role in the overall permeation behaviour. For instance, the external volume (enclosing the permeation cell) in the study by Calderoni *et al.* is intended to capture the radial loss of tritium flux from the downstream volume. As the study maintains, this volume is not purged during normal operations to allow for a build-up of tritium inventory, thus reducing the radial losses. It is, therefore, an underlying assumption of the study that the resultant partial pressure of tritium in the external volume has no significant impact on permeation behaviour in the downstream volume of the permeation cell.

It could, however, be argued that the flux leakage occurs from the upstream volume as well, where the tritium partial pressure is relatively higher. The axial configuration set-up in the study mandates a total surface area (including the bottom lid) of the upstream volume to be significantly larger than the Ni membrane cross-sectional area. Considering the permeability of FLiBe being orders of magnitude lower than the permeability of stainless steel [48-49], the transaction of flux between the upstream volume and external enclosing volume remains non-negligible. A concentration build-up over time will inevitably result in back permeation towards the vessel via path-III, as identified in **Fig.2**, leading to an overestimated FLiBe permeability and overshadowing the effect of possible oscillations from bubble releases. However, due to a lack of data on the temporal evolution of permeation flux in that study, a definite conclusion could not be drawn. In contrast, HYPERION experiments are designed to actively arrest any back permeation by real-time monitoring of $Q_2$ concentration in $V_{GB}$ and by consistently maintaining $P_{GB,Q2} < P_{down,Q2}$, as discussed under *Section-2.2*.

Further, the possibility of partial membrane coverage due to a higher surface tension of FLiBe remains a critical factor, potentially enhancing the diffusion of tritium as a function of the exposed membrane surface area. This could allow an alternative low-resistance permeation path, thereby inhibiting salt supersaturation and, therefore, any significant bubble nucleation and growth at the solid-liquid interface. As the salt coverage monitoring arrangements are not reported in earlier studies, these claims could not be verified. Moreover, the partial membrane coverage effect could also in part be assisted by a lower $P_{up,T2}$ used by Calderoni *et al.* A reduction in the source term could delay the salt supersaturation, and the impact of bubbles may go unnoticed over the experimental durations of interest.

For the radial permeation set-ups used by Anderl *et al.*, Nakamura *et al.* and Nishiumi *et al.*, the $Q_2$ permeation occurs via solid → liquid (salt) → solid, effectively presenting two solid-liquid interfaces before collection of the permeated $Q_2$ via sweep gas. Any oscillations in the permeation flux are therefore damped out by the second liquid-solid interface and may go unnoticed in an averaged permeation behaviour.

In addition to experiencing the dampening effect of multiple solid-liquid interfaces, vertically oriented radial permeation set-ups remain prone to the possibility of direct permeation (bypassing the salt altogether) between the innermost $Q_2$ tube and the outermost sweep gas collection tube via an arbitrary gas volume above the salt column. This situation becomes analogous to the partially covered membrane of axial permeation configurations. The implementation of permeation barriers in the top volumes of vertical annular cylindrical probes by Anderl *et al.* [47] is suggestive of such a possibility, although specific material details and/or efficiency (PRF) for these barriers remain unknown.

For horizontally oriented radial permeation tubular configurations [17-18], the high upstream $H_2$ pressure (up to 6 bar) could have likely resulted in a supersaturated state of FLiBe with the formation of bubbles due to the differential permeabilities of Monel-400 and FLiBe. It is worth mentioning that a horizontal tertiary tube arrangement in these studies provides a large interfacial surface area to allow the formation of multiple bubbles at the first solid-liquid interface. However, as discussed earlier, the second liquid-solid interface dampens any sharp signals at the measurement side and averages out the permeated concentration over time, leading to an overestimation of permeability. This claim is in view of a large difference observed for the estimated hydrogen permeabilities in FLiBe by the same group over a year (2015-2016) [17-18].

Another design ambiguity in the study by Nakamura *et al.* lies in the reported permeation area. The length of the $H_2$-pressurized innermost tube (= 750 mm) permeating to the salt over an annular tube length of 530 mm presents ambiguity in the correct accounting of the permeated $H_2$ concentration. This is due to the Ar collection stream that only sweeps the permeated $H_2$ over a length of 300 mm through the outermost tube. Such an arrangement could potentially allow an additional $H_2$ concentration build-up in the "unpurged" salt tube section ($\Delta l$ = 230 mm). This concentration could either support bubble formation at the solid-liquid interface or could diffuse to the salt section being purged by Ar ( = 300 mm). Accounting for a correct permeation area in such a configuration remains challenging. This could potentially be one of the reasons why the lengths for the salt-containing tube and the Ar sweep tube were revised to equal dimensions (= 530 mm) in the subsequent study by Nishiumi *et al.* (2016).

*3.4.3 Impact of simultaneous presence of multi-component hydrogen isotopes*

The presence of $Q_2$ (Q ≡ H/D) on the downstream side of the permeating membrane, as employed by Calderoni *et al.*, is argued to assist the overall tritium permeation through isotopic exchange, owing to a higher diffusivity exhibited by HT and DT [21]. This claim is also supported through the conclusions drawn by Calderoni *et al.*, asserting the possible diffusion of tritium as HT, instead of $T_2$, within bulk FLiBe. Therefore, an accelerated tritium diffusion resulting from a hydrogen-saturated downstream volume is expected to suppress the probability of significant bubble nucleation. Moreover, as highlighted by Carotti *et al.* (2021), the solubility values and associated activation energies in the study by Calderoni *et al.* present contradictory evidence towards establishing the chemical form of tritium within FLiBe [50].

To summarize, an intermittent transport through bubbles over long operational durations and continuous direct transport via cover gas regions could lead to an overestimated salt permeability in the axial and radial permeation configurations. In essence, although the bubble nucleation phenomenon at the solid-liquid interface is expected to significantly govern the $Q_2$ transport behaviour in both axial and radial permeation experiments, systematic investigations - numerical or experimental - exploring this aspect are severely lacking.

Surface roughness is also known to assist the bubble nucleation process [51-52]. To explore and possibly resolve the impact of surface roughness-induced interface bubble nucleation on the $Q_2$ transport in HYPERION, systematic checks are performed by charging $Q_2$ (Q ≡ H/D) at the salt side. This configuration is similar to that adopted by Zeng *et al.* (2019) for permeation studies in molten FLiNaK [53]. The results obtained from the revised charging methodology are discussed under *Section-3.5*.

### 3.5. Permeation experiments for FLiBe: $Q_2$ charging at the salt side

**Fig.9** presents the measured hydrogen permeation flux for the salt-side charging configuration. The magnitude of $J_{measured}$ is observed to increase by a factor of ~ 3 compared to that in the apparent steady state for metal-side charging at 873 K (**Fig.7**). Moreover, clearly distinct permeation flux values are observed for $H_2$ and $D_2$, as shown in **Fig.10**, in stark contrast to the observations made in **Fig.8**. A higher magnitude of flux for $H_2$ as compared to that for $D_2$ suggests a higher permeability for $H_2$. This establishes an isotopic permeation behaviour rather than an interface-limited transport in the revised charging methodology. These observations strengthen the claim that, in the present study, the $Q_2$ transport for metal-side charging is impeded primarily by the interfacial bubble nucleation, while impurities play a rather insignificant role. These results further establish that $Q_2$ charging at the low permeance (FLiBe) side could help suppress the permeation barrier effect associated with the solid-liquid interface.

However, after peaking, a gradual decrease in the steady state flux at higher temperatures (≥ 873 K) still suggests a non-negligible effect of gradual bubble nucleation and growth over time. From these initial observations, the present understanding of bubble nucleation and growth in the HYPERION facility is depicted in **Fig.11(a-d)**. An ideal wetting scenario (**Fig.11(a)**) provides a series-coupled interface leading to $J_{Ni} = J_{FLiBe}$, with no sinks at the interface. This condition is analogous to an electrical circuit with two series-connected resistors, providing an overall higher resistance to modulate the total current in the circuit. The driven current (~ permeation flux) remains the same through both resistors connected in series. In contrast, surface roughness and nano/micro-bubbles resulting from a non-ideal interface (**Fig.11(b)**) can behave like capacitor(s) at the bubble sites, analogous to a multi-branch parallel electrical circuit, by providing potential trapping sites for $H_2$ and $Ar + H_2$. In the present study, $L_{Ni} < L_{FLiBe}$ while $Ø_{Ni} >> Ø_{FLiBe}$. Therefore, as shown in **Fig.11(c)**, $Q_2$ charging at the Ni side renders a much higher $P_{interface,Q2}$ driving rapid bubble nucleation and growth at the interface, while the downstream FLiBe layer restricts the $Q_2$ permeation across it. In this configuration, the effective area available for permeation depletes at a faster rate due to rapid bubble growth, leading to (i) an overall reduced $J_{measured}$ (**Fig.6**), (ii) frequent bubble releases under the influence of buoyancy (**Fig.7**), and (iii) masking of observable isotopic permeation effects (**Fig.8**). This state invalidates the application of eq. (7) and (10), as the flux across both materials does not remain the same. These first insights from the HYPERION investigations could provide a possible explanation for mechanisms that could lead to discrepancies in the observed hydrogen isotope permeation effects. It is worth mentioning that the possibility of permeant charging at the salt side in radial configurations is constrained by liquid layer containment requirements. Therefore, the radial permeation configurations remain inherently prone to interface bubble effects.

The $Q_2$ charging at the FLiBe side, however, effectively reduces $P_{interface,Q2}$ available to support the interface bubble growth due to a large $Q_2$ pressure drop across the low-permeability salt layer (**Fig.11(d)**). The sink effect of the Ni-FLiBe interface is further significantly reduced owing to a higher permeability Ni membrane available downstream, prohibiting any significant $Q_2$ concentration build-up. This configuration leads to a much-delayed interface bubble nucleation, as observed through a gradual decrease in the $H_2$ permeation flux at 873 K and 973 K in **Fig.9**. Virtually no decrease observed in the $H_2$ flux at 773 K (**Fig.9**), and for $D_2$ flux over the complete temperature range (**Fig.10**) clearly signifies reduced diffusivity for FLiBe at lower temperatures and for the heavier isotope, respectively.

Here, it is worth highlighting that in similar attempts with a 30 mm FLiNaK layer placed in an axial permeation set-up, Zeng et al. (2019) observed no difference in the permeabilities for $H_2$ and $D_2$ over a temperature range of 773 K – 973 K. This indicates the possibility of significant membrane area coverage by interface bubbles masking the isotopic effect, as observed during metal-side charging for the present HYPERION study (**Fig.8**).

### 3.6. Permeability estimations for FLiBe: $H_2$ charging at the salt side with improved wetting scenarios

To observe the effect of enhanced interface wettability on hydrogen transport, $H_2$ charging is preceded by vessel shaking performed over 4 h at regular intervals of 15 min with a shaking duration of 60 s each. As depicted in **Fig.12**, improvements between 11% and 36% are observed in the magnitude of $J_{measured}$ compared to that in **Fig.9** over a temperature range of 773 K to 873 K. Further, the gradual decrease observed at 873 K (**Fig.9**) seems arrested, likely due to a reduction in the population of interface nucleation sites through shaking-induced wetting enhancements.

Even in this situation, the permeation experiments at temperatures above 873 K clearly depicted a significant impact of bubbles on the magnitude of the permeation flux. A continuous flux decrease could only be compensated through regular vessel shaking over the entire duration of the experiments at 923 K and 973 K. Further, the observed flux at 973 K remains lower than that at 923 K. This provides a lower confidence in the estimated FLiBe permeability values for temperatures above 873 K from current investigations in the HYPERION facility.

**Fig.13(a)** provides a comparison among the estimated hydrogen permeabilities (salt-side charging), interface-limited apparent permeability (metal-side charging), and permeabilities estimated for hydrogen isotopes from earlier works over the range of 773 K - 973 K. Comparisons of the data for metal-side charging and salt-side charging configurations in HYPERION suggest that the presence of bubbles could suppress the permeability values by ~ 77% over a temperature range of 773 K - 873 K. The interface-limited regime (metal-side charging) reduces the apparent FLiBe permeability by a factor of ~ 3.3 at 773 K and ~ 3.9 at 873 K, although this factor reduces to ~ 1.6 at 973 K. A convergence of the estimated permeabilities for the two charging configurations suggests a significant bubble impact at temperatures above 873 K for either charging methodology. In this case, the calculated activation energies are 26.8 kJ/mol for the PRF = ∞ case and 29.1 kJ/mol for the PRF = 1 case for the salt-side charging, considering the full range up to 973 K. Further detailed experiments are necessary to gain a better understanding of these coupled effects at the Ni-FLiBe interface.

Considering the limitations arising from the presence of bubbles in this study, a higher-confidence interval is taken as 773 K - 873 K for permeability estimations of $H_2$ in FLiBe, where the bubble impacts are assessed to be relatively less severe (**Fig.12**). In contrast, the $D_2$ permeability in FLiBe is estimated up to 973 K, considering a much lower bubble impact (**Fig.10**). Estimates of $Ø_{D2-FLiBe}$ are performed as per eq. (11) and (12), where the required values for $Ø_{D-Ni}$ are derived from $Ø_{H-Ni}$ using a factor of $\sqrt{2}$ in accordance with the classical theory for isotopic permeation. Urrestizala et al. (2024) provided a rigorous experimental validation for the applicability of this factor for H/D permeabilities in SS-316 [54]. Further, Noh et al. (2014) evaluated this factor to lie between 1.4 and 1.6 from several experimental studies on H/D permeation in Ni [39]. The ratio of $Ø_{D-Ni}$ evaluated using this approach to the corresponding deuterium permeabilities in Ni from earlier works (**Fig.4**; ref [37-39]) lies within a range of 0.75 – 1.30.

**Fig.13(b)** provides a consolidated comparison of the higher-confidence $Q_2$ permeabilities in FLiBe estimated from HYPERION experiments. The combined percentage uncertainty, derived from the root-sum-square of individual uncertainties, remains within ± 41%. For comparison purposes, results from a few studies with FLiNaK are also included in **Fig.13(a,b)**.

The intrinsic permeability of $H_2$ in FLiBe (units: $H_2/m.s.Pa$) over the range 773 K – 873 K is correlated to the temperature as per the following Arrhenius equations:

$$Ø_{H2-FLiBe(PRF=1)} = 5.97 \times 10^{13} \exp(-51.3 \times 10^3/R_gT)$$

$$Ø_{H2-FLiBe(PRF=\infty)} = 3.81 \times 10^{13} \exp(-48.7 \times 10^3/R_gT)$$

The intrinsic permeability of $D_2$ in FLiBe (units: $D_2$/m.s.Pa) over the range 773 K – 973 K is correlated to the temperature as per the following Arrhenius equations:

$$\emptyset_{D2-FLiBe(PRF=1)} = 7.97 \times 10^{12} \exp(-44.2 \times 10^3/R_gT)$$

$$\emptyset_{D2-FLiBe(PRF=\infty)} = 6.32 \times 10^{12} \exp(-42.8 \times 10^3/R_gT)$$

A difference of ~ 7 kJ/mol observed in the activation energies between the two isotopes suggests a higher temperature sensitivity of $H_2$ permeability in FLiBe. The experimentally estimated permeability of $H_2$ in FLiBe remains higher than that of $D_2$ in FLiBe by a factor of 2.3 to 2.4 in the temperature range 773 K – 873 K.

It is worth noting that the estimated $H_2$ permeability at 773 K converges well with the values reported by Nakamura *et al.* and Nishiumi *et al.* However, at 873 K, the estimated permeability from HYPERION is a factor of ~ 3 lower than that reported by Nakamura *et al.*, and an order of magnitude lower than that by Nishiumi *et al.* For radial permeation set-ups, as Lam argues [20], a simplified analytical model for the cylindrical diffusion geometry, and selection of hydrogen-interacting Monel-400 may lead to errors in the estimated transport parameters. It could be reasoned that at higher temperatures, both these factors have a greater impact, leading to a much higher slope compared to all other permeation studies with FLiBe (**Fig.13(b)**). Further, as the temperature increases, the probability of interface bubble nucleation is significantly higher in the radial permeation set-ups due to a directed $Q_2$ permeation from high permeability metal to low permeability salt, similar to the metal-side charging configuration in HYPERION. As argued under *Section 3.4.2,* the second liquid-solid interface aids in the dampening of sudden increases in downstream permeant concentration, but can provide an overall higher averaged permeation flux, resulting in an overestimated salt permeability.

The tritium permeability in FLiBe reported by Calderoni *et al.* remains much higher than the corresponding hydrogen permeabilities reported by Nakamura *et al.* and Nishiumi *et al.* over a wide range of temperature values. A similar observation can be made for the estimated hydrogen permeability in FLiNaK by Zeng *et al.*, compared to that by Nakamura *et al.* Relevant possible design choice factors giving rise to such counter-intuitive observations are discussed under *Sections 3.4.2 and 3.4.3*. Moreover, the susceptibility to a convection-assisted transport is higher in a relatively thicker salt layer (FLiBe layer: 8.1 mm - 13.7 mm in Calderoni *et al.*, FLiNak layer: 30 mm in Zeng *et al.*, and FLiBe column ~ 150 mm in Anderl *et al.*). Hailin (2017) experimentally demonstrated an order of magnitude increase in the deuterium permeation flux due to natural convection in a 15 mm thick GaInSn layer at 450˚C [55]. This could inflate the estimated permeability by providing an alternative, lower resistance mode of transport, which is not related to diffusion. In contrast, the thicknesses of FLiBe layers used in Nakamura *et al.*, Nishiumi *et al.* and HYPERION experiments are less than 6 mm.

Similarly, in addition to the factors discussed under *Section 3.4*, the simplifying assumptions for analytical solutions in previous works are expected to impact the estimated transport parameters. For instance, Nishiumi *et al.* neglect the $H_2$ diffusion through Monel-400, effectively coupling the overall effect in the transport parameters for FLiBe. Such assumptions are explicitly avoided in the data analysis for HYPERION by accounting for a finite permeation resistance offered by the Ni membrane. The experimental data from dry runs are used to calculate the Ni-FLiBe interface pressure ($P_{interface,H2}$) as per eq. (7), (10) and (12), to arrive at the intrinsic permeability of $Q_2$ in FLiBe.

The distinct isotopic effects observed for $H_2$ and $D_2$ for the two charging configurations in the present study (**Fig.8** and **Fig.10**) provide initial insights into the possible processes leading to conflicting observations from earlier works [53]. However, more refined investigations are further needed to quantify the permeability ratios for different hydrogen isotopes in FLiBe at higher temperatures.

## 4. CONCLUSIONS

Systematic experimental efforts are made to quantify the permeabilities of hydrogen and deuterium in molten FLiBe over a temperature range of 773 K – 973 K, relevant to applications in next-generation fusion breeding blankets. The HYPERION facility is designed with salient features to reduce the experimental uncertainties and to eliminate unsubstantiated assumptions for derived transport parameters. These investigations experimentally explore - for the first time - the effect of interface-limited transport in FLiBe, suggesting bubble formation as a transport barrier mechanism reducing permeation by ~77%. The proposed hypotheses are validated using the differential response of the thermal conductivity detector towards $H_2$ and $D_2$ permeation in diffusion-limited and interface-limited transport regimes. Potential sources of error in the previous studies are identified and discussed in detail. Results from the present study help inform practical constraints on the charging methodology for future efforts to quantify intrinsic permeabilities and isotopic effects. Further upgrades are planned for the HYPERION facility to enable diffusivity and solubility estimations for FLiBe and other candidate salts. In particular, the focus areas for future investigations will include visual monitoring and behaviour characterization of bubbles at the Ni-FLiBe interface, systematic remedial measures to decouple the impact of bubbles from estimated permeabilities, exhaustive characterization of the isotopic dependence relevant to tritium transport database generation, studies to explore the impact of controlled salt redox variations on the overall transport behaviour, and efforts to expand hydrogen transport modeling capabilities for 2D and 3D permeation scenarios.

## 5. ACKNOWLEDGEMENTS

This work was supported by RPP039: FLiBe salt compositional analysis and its impact on hydrogen transport and corrosion, from Commonwealth Fusion Systems (CFS). Partial support for this research was also provided by the core center grant P30-ES002109 from the National Institute of Environmental Health

Sciences, which supported the use of the digestion system and ICP-MS at the MIT Center for Environmental Health Sciences.

The authors gratefully acknowledge the active support of Rick Leccacorvi and Rui Vieira from MIT PSFC, as well as the staff from the MIT Central Machine Shop, for their assistance in fabricating the permeation cell. The authors also appreciate technical discussions with Nikola Goles during the initial phase of facility setup.

## 6. Data Availability Statement

The data that support the findings of this study are available in the open data repository at:

https://doi.org/10.5281/zenodo.19240266.

## 7. Declaration of Competing Interests

The authors declare the following financial interests/person relationships, which may be considered as potential competing interests: financial support was provided by Commonwealth Fusion Systems (CFS). The authors declare that there are no other conflicts of interest, financial or otherwise, that could have influenced the results of this study.## 8. CRediT authorship contribution statement

**A. Saraswat:** Conceptualization, Methodology, Software, Validation, Formal analysis, Investigation, Data curation, Writing – Original Draft, Visualization. **W. Zhou:** Conceptualization, Methodology, Validation, Resources, Writing – Review & Editing, Visualization, Supervision. **N. Kim:** Investigation, Writing – Review & Editing. **J. F. Cota:** Methodology. **G. Zheng:** Writing – Review & Editing, Supervision. **A. A. Khechfe:** Writing – Review & Editing, Supervision. **C. S. Barthel:** Writing – Review & Editing, Supervision, Funding acquisition. **M. P. Short:** Methodology, Writing - Review & Editing, Supervision. **R. Delaporte-Mathurin:** Methodology, Writing - Review & Editing, Supervision. **K. B. Woller:** Conceptualization, Methodology, Resources, Supervision, Funding acquisition.

## REFERENCES

[1] A. J. Barker et al., Analysis of a segmentation approach to breeder blanket design and the utilisation of FLiBe as a novel neutron reflector, Nuclear Fusion, 65(8), 2025, 086022.

[2] L. Zhang et al., Liquid metals power advanced nuclear energy systems, The Innovation, 6(9), 2025, 100959.

[3] S. Smolentsev et al., Overview of magnetohydrodynamic studies for liquid metal systems of a fusion power reactor at Oak Ridge National Laboratory, Nuclear Science and Engineering, 2025, 1-16.

[4] R. Nishio et al., Experimental and analytical investigations to reduce MHD pressure drop for liquid LiPb fusion blanket systems: use of ODS-FeCrAl alloys with electrically insulating α-$Al_2O_3$ layer in optimal flow channel geometry, Nuclear Materials and Energy, 44, 2025, 101965.

[5] A. Saraswat et al., Experimental investigations on electrical-insulation performance of $Al_2O_3$ coatings for high temperature PbLi liquid metal applications, Annals of Nuclear Energy, 167, 2022, 108856.

[6] R. Boullon et al., Molten salt breeding blanket: Investigations and proposals of pre-conceptual design options for testing in DEMO, Fusion Engineering and Design, 171, 2021, 112707.

[7] S. Segantin et al., Optimization of tritium breeding ratio in ARC reactor, Fusion Engineering and Design, 154, 2020, 111531.

[8] Y. Zhu and A.I. Hawari, Thermal neutron scattering cross section of liquid FLiBe, Progress in Nuclear Energy, 101, Part C, 2017, 468-475.

[9] B. Sorbom et al., ARC: A compact, high-field, fusion nuclear science facility and demonstration power plant with demountable magnets, Fusion Engineering and Design, 100, 2015, 378-405.

[10] S. Ferry et al., The LIBRA experiment: Investigating robust tritium accountancy in molten FLiBe exposed to a D-T fusion neutron spectrum, Fusion Science and Technology, 79(1), 2022, 13–35.

[11] H. Yang et al., Validation of FESTIM hydrogen transport modeling in FLiBe through HYPERION permeation data, IAEA Workshop on Digital Engineering for Fusion Energy Research, 2025.

[12] K. Baral et al., Temperature-dependent properties of molten $Li_2BeF_4$ salt using *Ab Initio* molecular dynamics, ACS Omega, 6(30), 2021, 19822-19835.

[13] R. D. Scheele et al., A laboratory-scale process for producing dilithium beryllium tetrafluoride (FLiBe) with dissolved uranium tetrafluoride, Journal of Nuclear Materials, 585, 2023, 154636.

[14] B. Lu et al., Progress on tritium toxicity and detoxification, ACS Chemical Health & Safety, 31(2), 2024, 144-152.

[15] V. Dolin et al., The impact of tritium in the environment, Applied Sciences, 15(12), 2025, 6664.

[16] A. P. Malinauskas and D. M. Richardson, The solubilities of hydrogen, deuterium, and helium in molten $Li_2BeF_4$, Industrial & Engineering Chemistry Fundamentals, 13(3), 1974, 242-245.

[17] A. Nakamura et al., Hydrogen isotopes permeation in a fluoride molten salt for nuclear fusion blanket, Journal of Plasma and Fusion Research SERIES, 11, 2015, 25-29.

[18] R. Nishiumi et al., Hydrogen permeation through Flinabe fluoride molten salts for blanket candidates, Fusion Engineering and Design, 109–111, Part B, 2016, 1663-1668.

[19] P. Calderoni et al., Measurement of tritium permeation in flibe (2LiF–BeF2), Fusion Engineering and Design, 83(7–9), 2008, 1331-1334.

[20] S. T. Lam et al., The impact of hydrogen valence on its bonding and transport in molten fluoride salts, Journal of Materials Chemistry A, 9, 2021, 1784-1794.
10

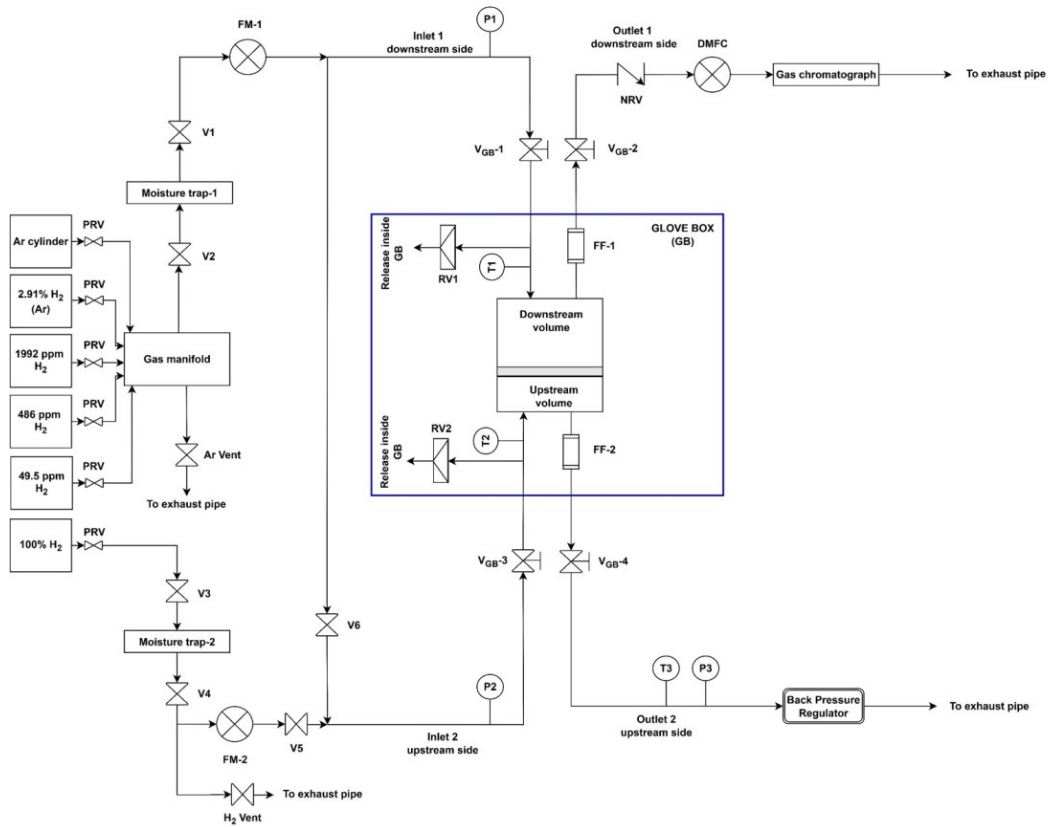

Fig.1(a) Process flow diagram for HYPERION facility (PRV: Pressure regulating valve, FM: Flowmeter, RV: Relief valve, FF: Filter, NRV: Non-return valve, DMFC: Digital mass flow controller)

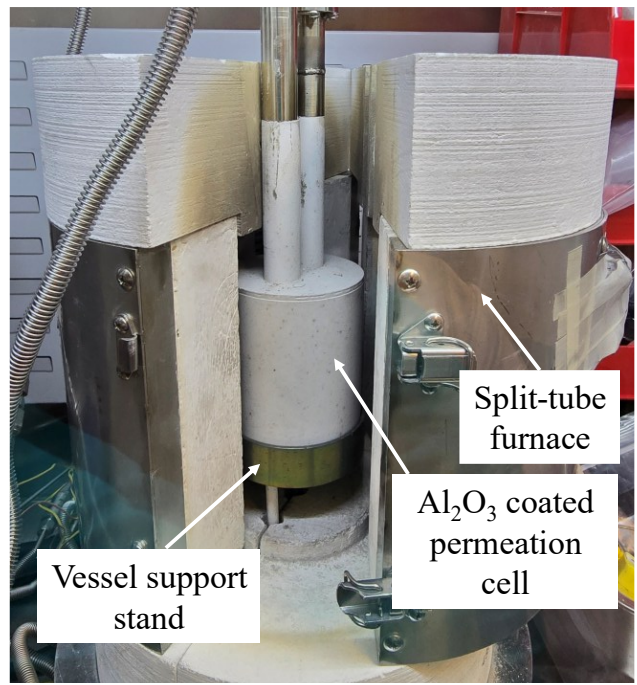

Fig.1(b) Permeation cell installed in a vertical split-tube furnace

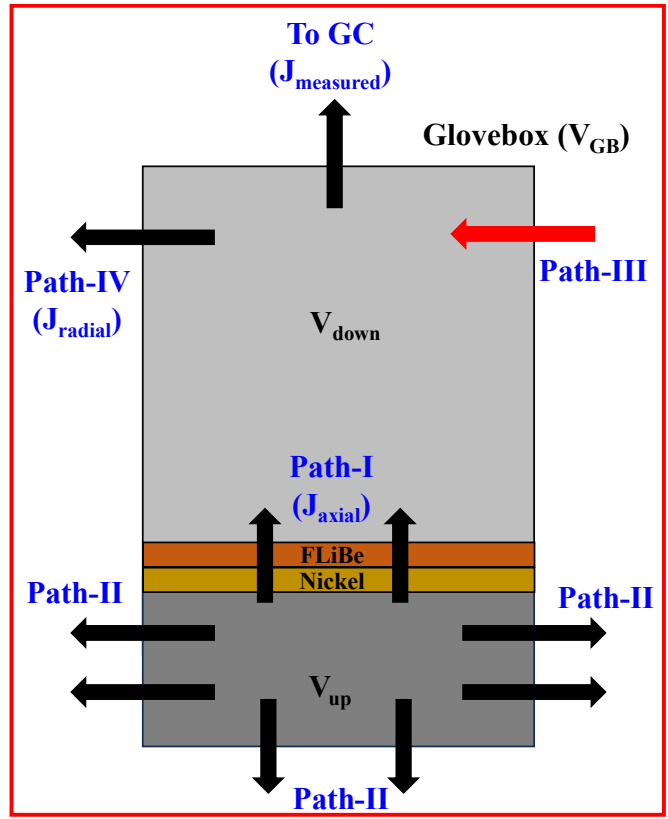

Fig.2 Possible permeation paths for the HYPERION set-up with $Q_2$ charging at the Ni side

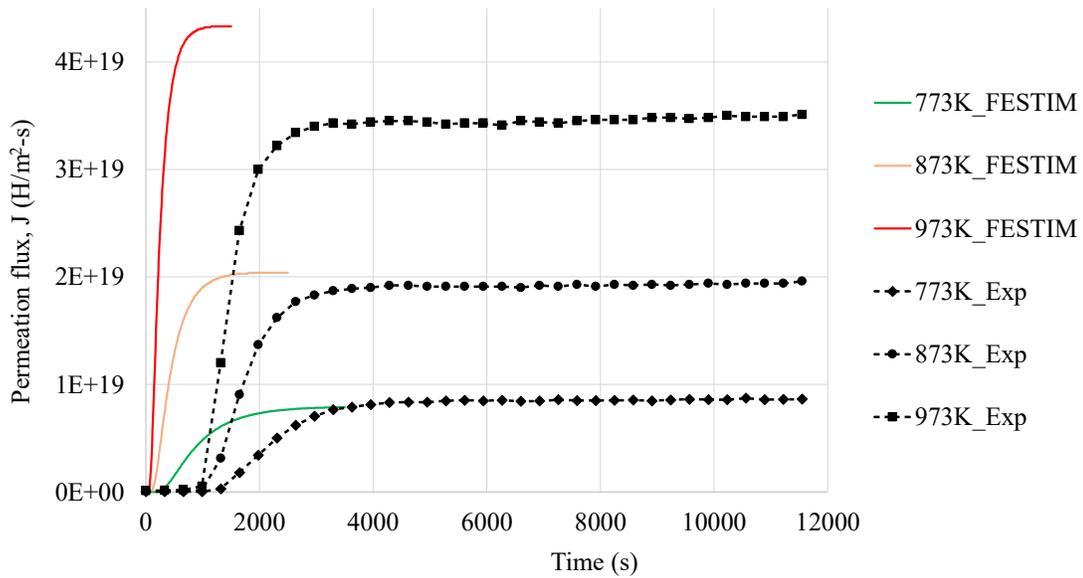

Fig.3 Comparisons of experimentally measured permeation flux and numerically estimated 1D permeation flux for hydrogen in Ni-200 over a temperature range of 773 K - 973 K (dry runs with no salt) at $P_{up,H2}$ of 0.11 MPa(a) (for FESTIM simulations, the solubility and diffusivity parameters are taken from [32])

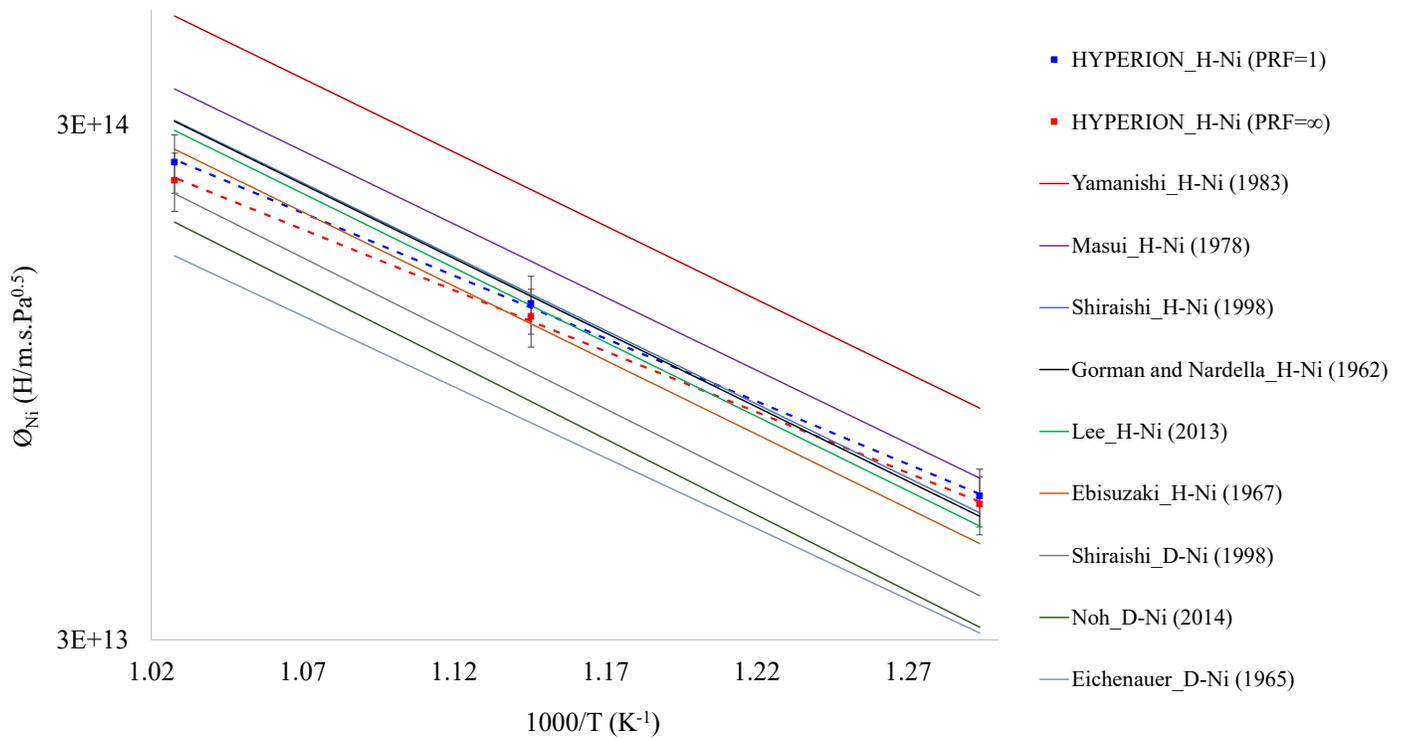

Fig.4 Arrhenius plots for hydrogen permeability in Ni-200 from 773 K to 973 K
(For interpretation of the color-coded data in the legend, please refer to the online version of this article)

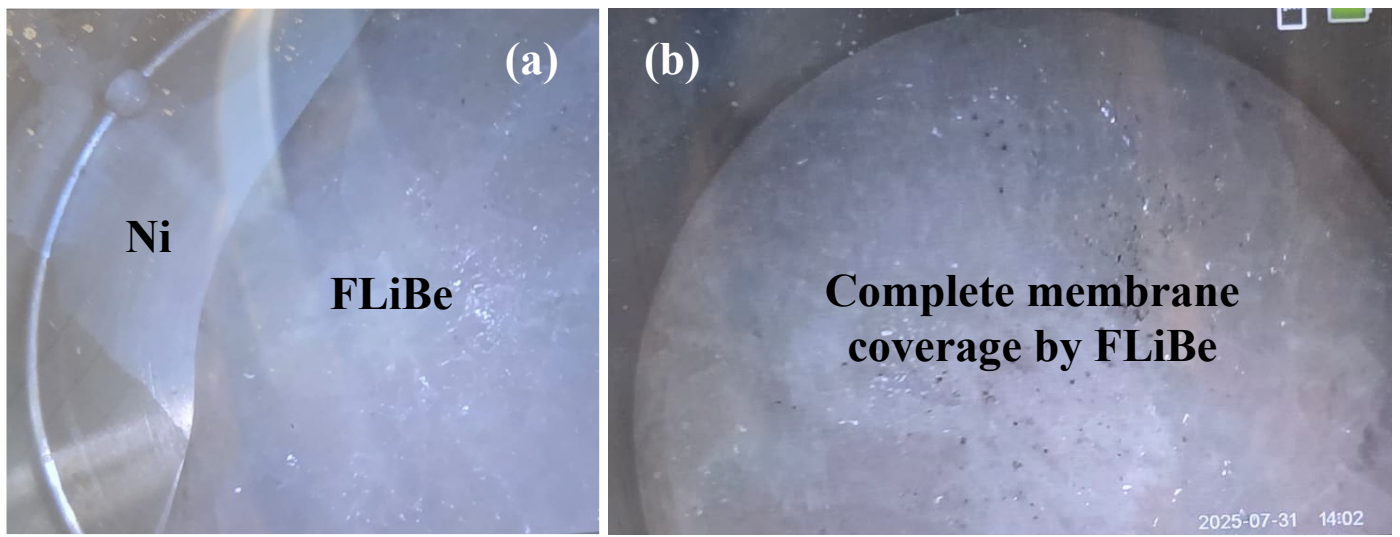

Fig.5 (a) Partial coverage of the Ni membrane by FLiBe; (b) Complete coverage achieved with the addition of more FLiBe inventory (membrane diameter = 77.98 mm)

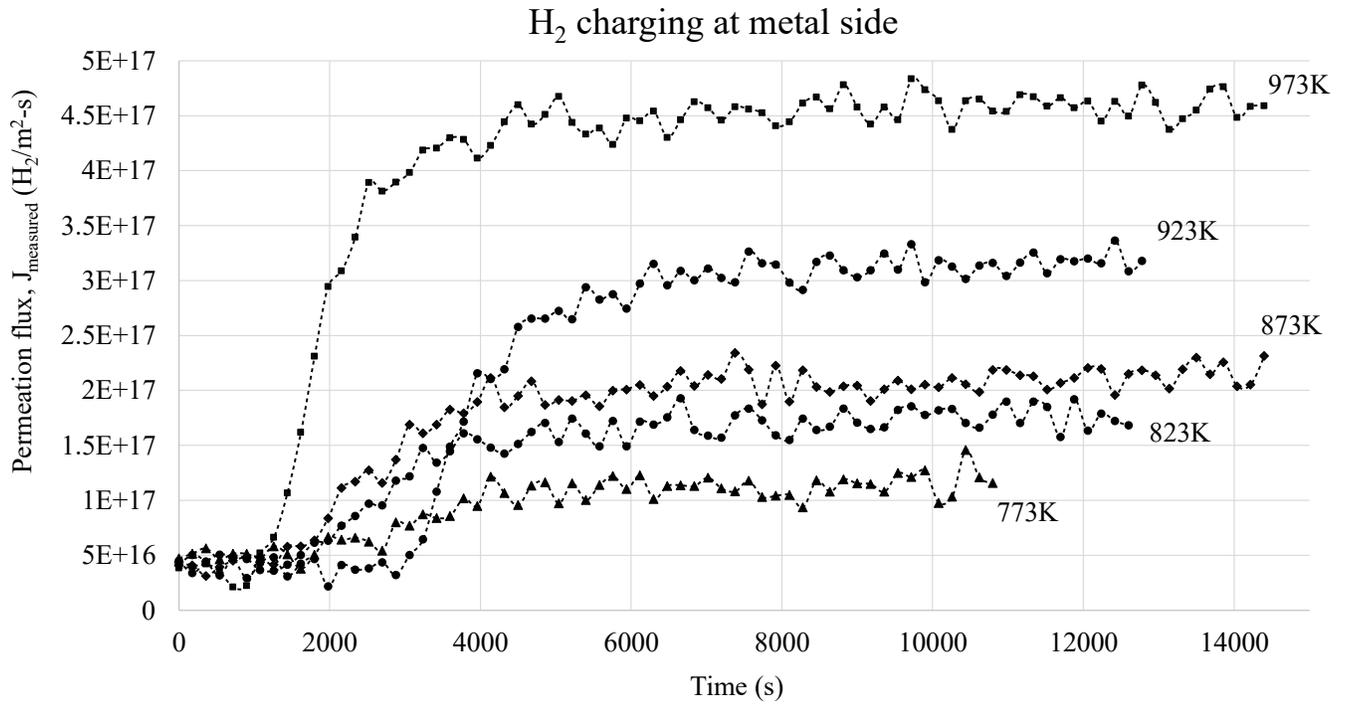

Fig.6 Experimentally measured hydrogen permeation flux for FLiBe at $P_{up,H2}$ of 0.10 MPa(a) with $H_2$ charging at Ni side

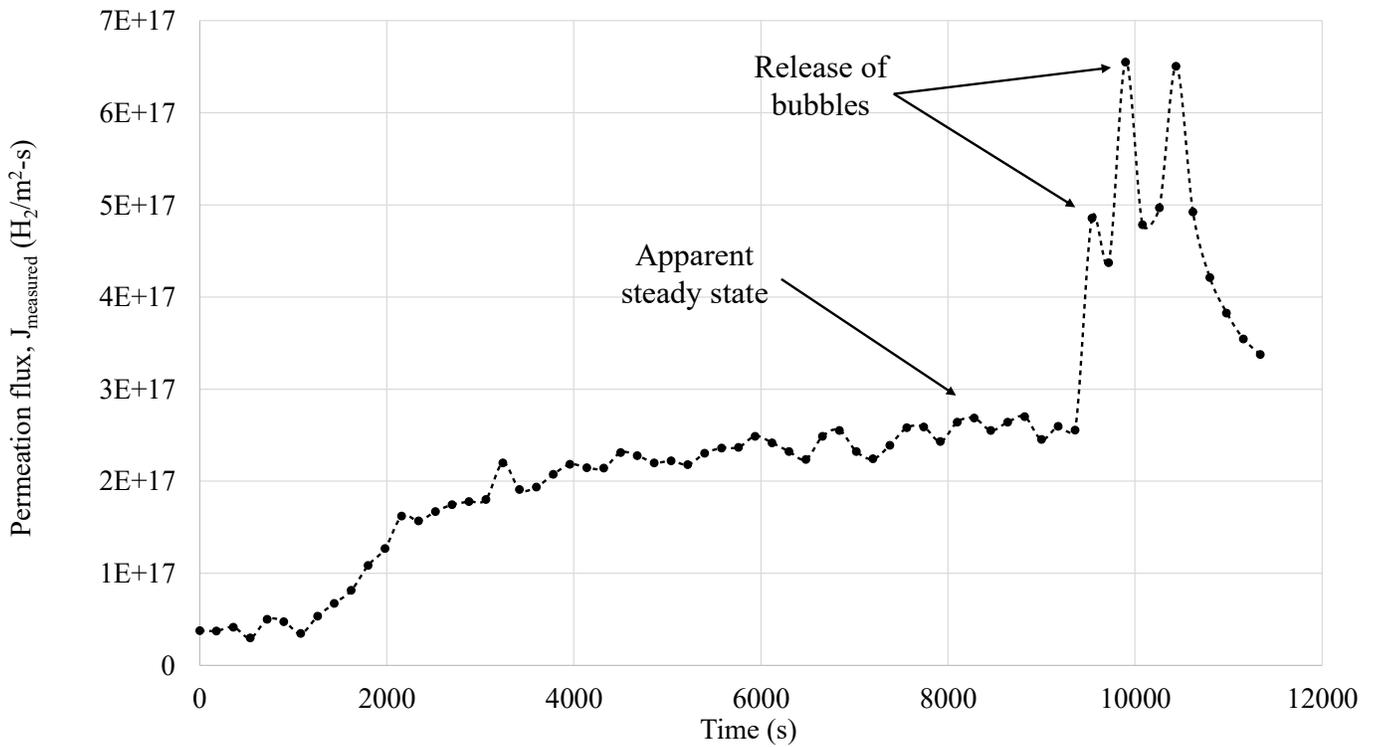

Fig.7 Oscillations in $J_{measured}$ indicative of bubble formation and detachment at Ni-FLiBe interface (T = 873 K, $P_{up,H2}$ = 0.13 MPa(a))

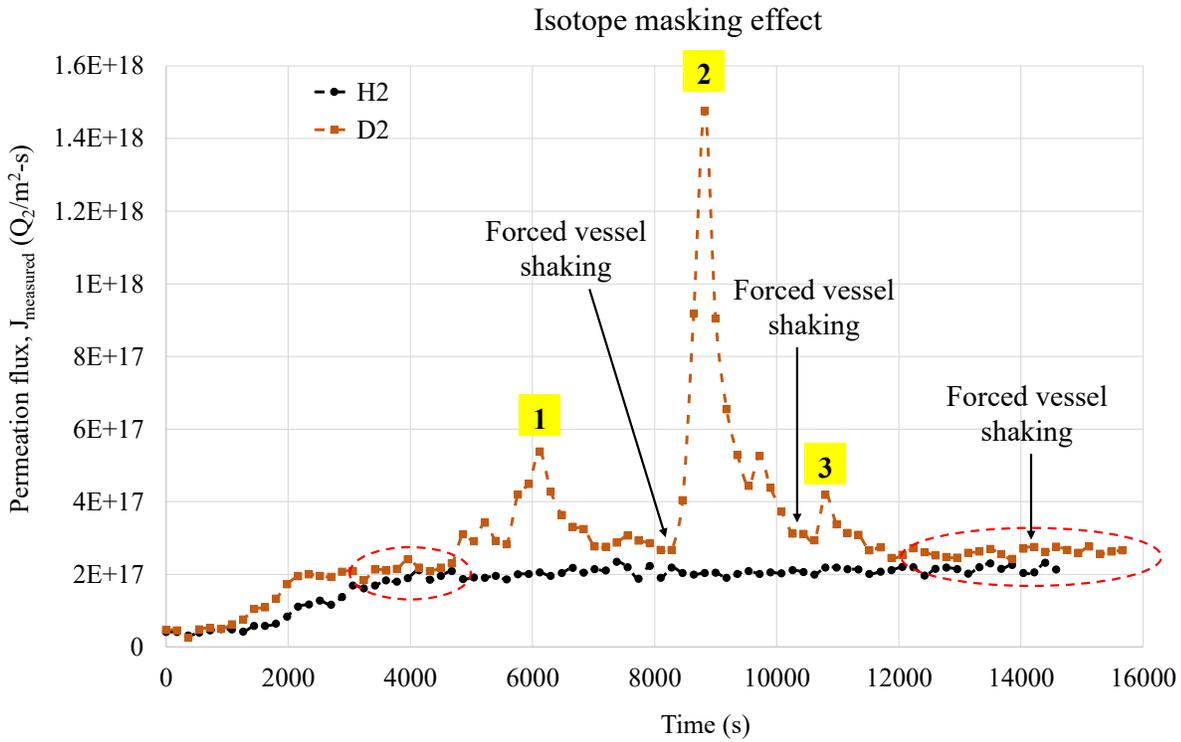

Fig.8 Comparison of permeation flux for $H_2$ and $D_2$, suggesting a substantial permeation barrier effect exhibited by the bubbles at the Ni-FLiBe interface (T = 873 K, $P_{up,Q2}$ = 0.10 MPa(a)); Region 1: Spontaneous detachment of bubbles; Region-2 and 3: Forced bubble detachment through manual vessel shaking

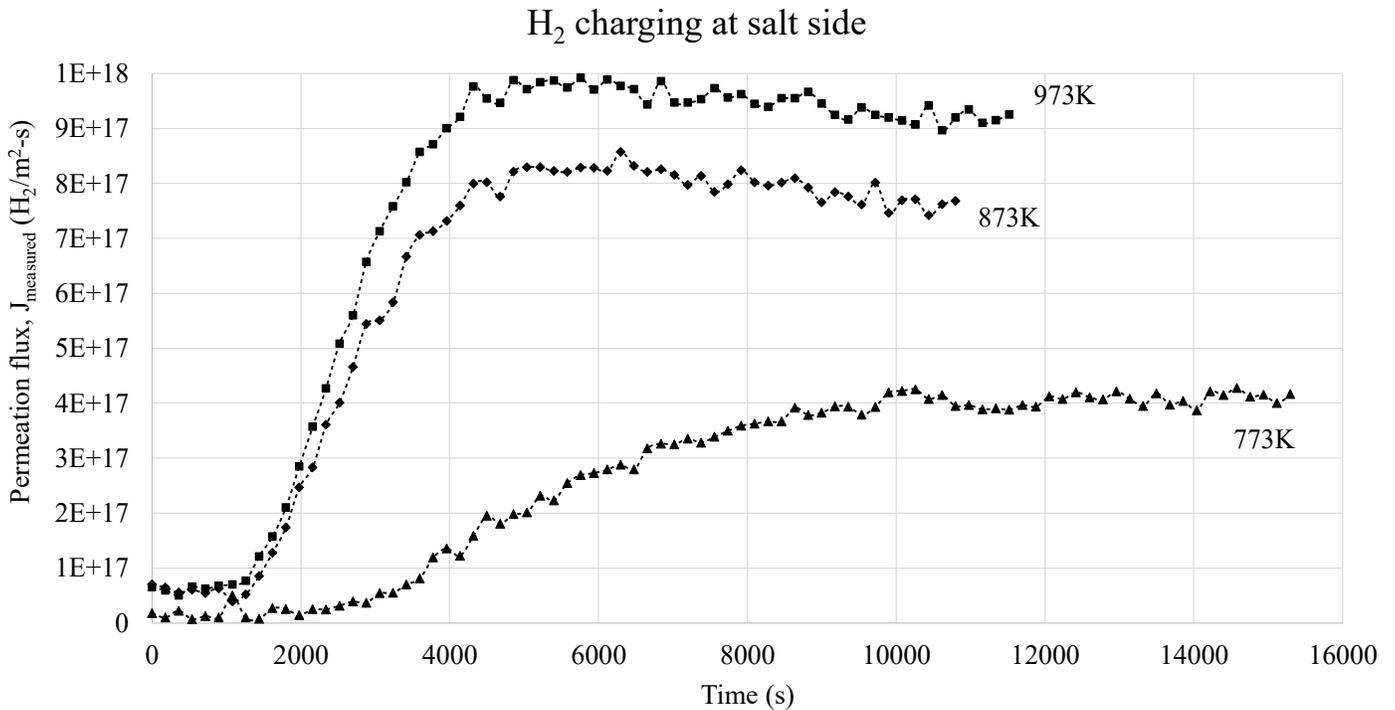

Fig.9 Experimentally measured hydrogen permeation flux for FLiBe at $P_{up,H2}$ of 0.13 MPa(a) with $H_2$ charging at FLiBe side

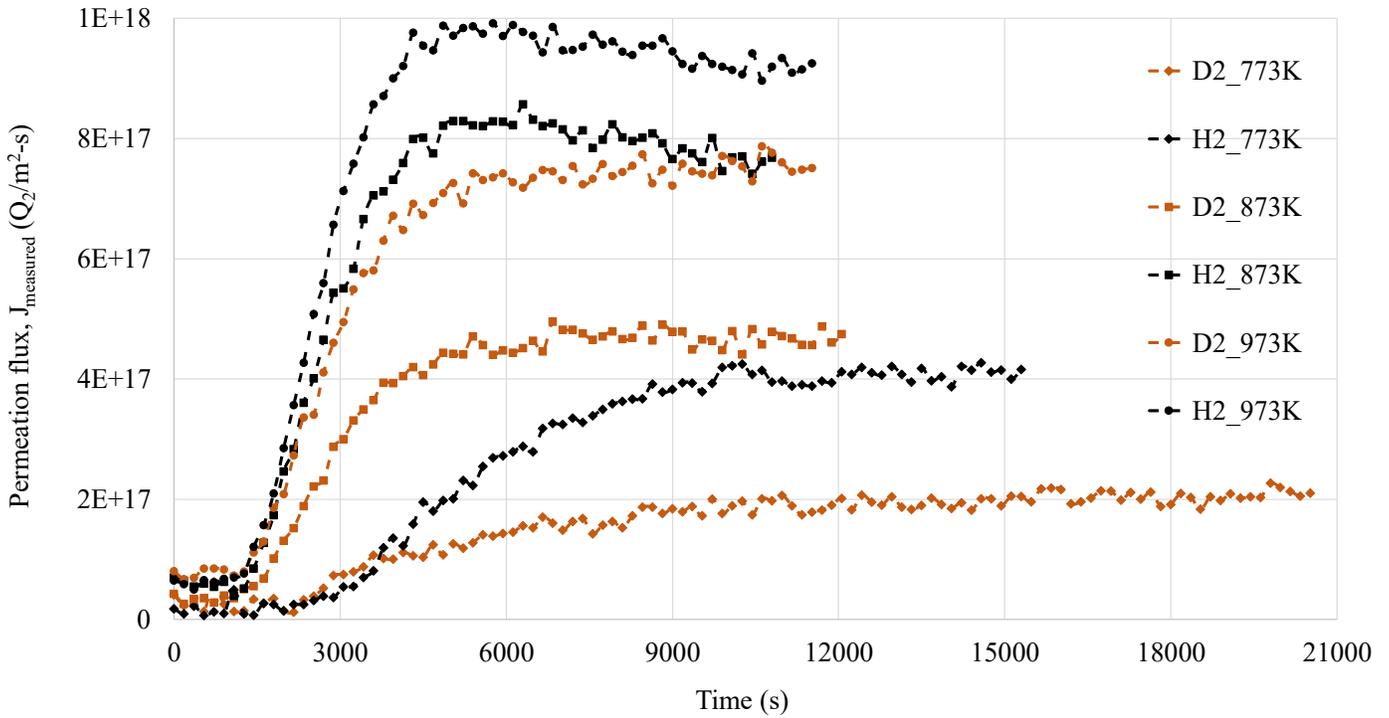

Fig.10 Comparisons of permeation flux for $H_2$ and $D_2$ (773 K - 973 K), suggesting an isotopic permeation effect with suppressed interface-limited transport for $Q_2$ charging at FLiBe side ($P_{up,Q2}$ = 0.13 MPa(a))

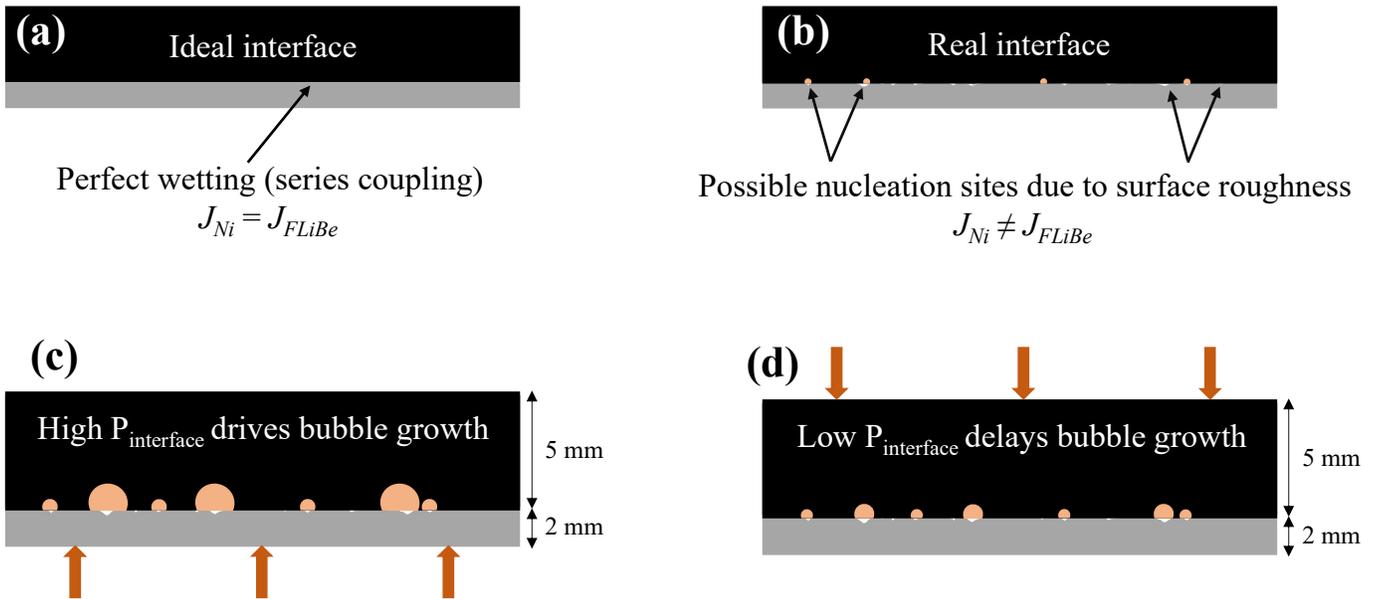

Fig.11 Schematic representation of bubble nucleation and growth at the Ni-FLiBe interface: (a) ideal coupling at a perfectly wetted solid-liquid interface; (b) imperfect wetting due to pre-existing micro-scale gas bubbles resulting from a coupled effect of finite surface roughness and higher surface tension of FLiBe; (c) bubble formation and growth for metal-side $Q_2$ charging; (d) bubble formation and growth for salt-side $Q_2$ charging

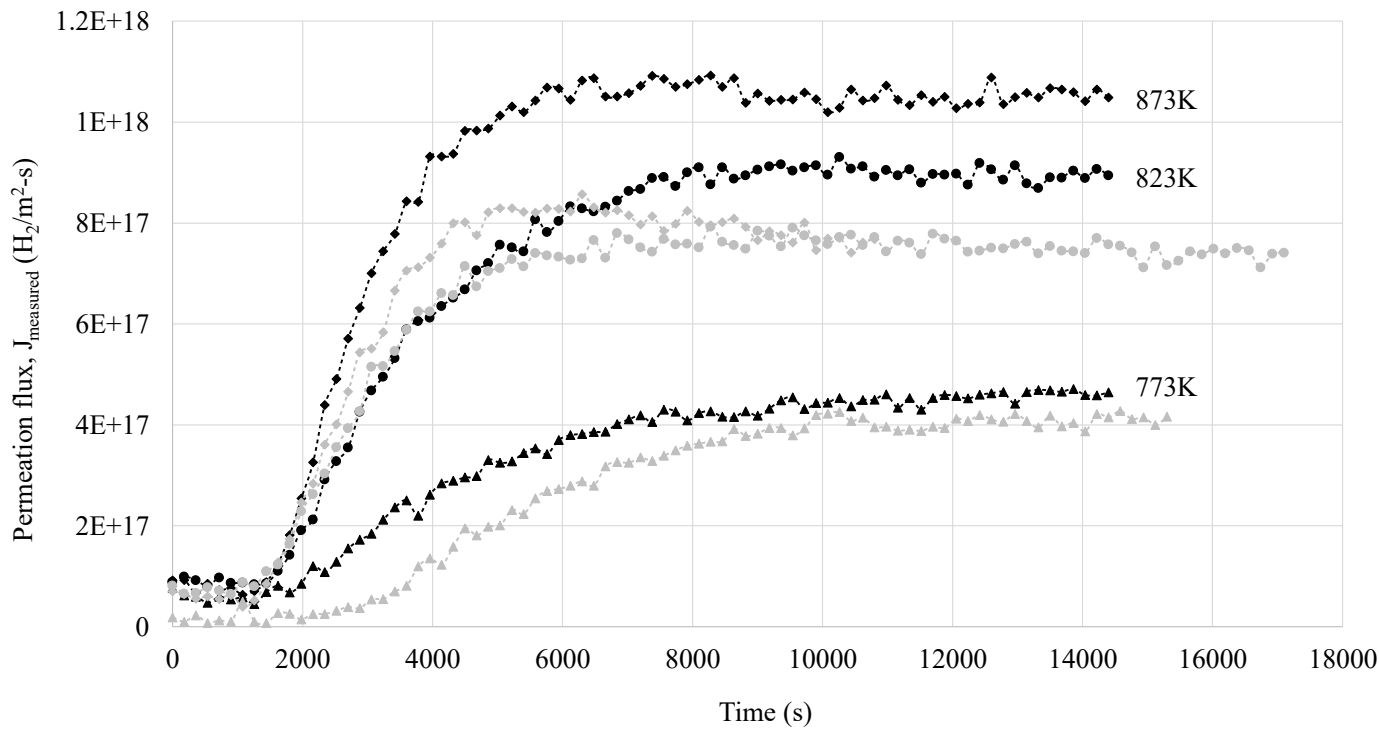

Fig.12 Increase in experimentally measured hydrogen permeation flux values for FLiBe at $P_{up,H2}$ of 0.13 MPa(a) with $H_2$ charging at FLiBe side after vessel shaking to enhance interface wettability (for comparisons, hydrogen permeation flux values without shaking-induced wettability enhancements are shown in grey color with markers corresponding to the same temperature)

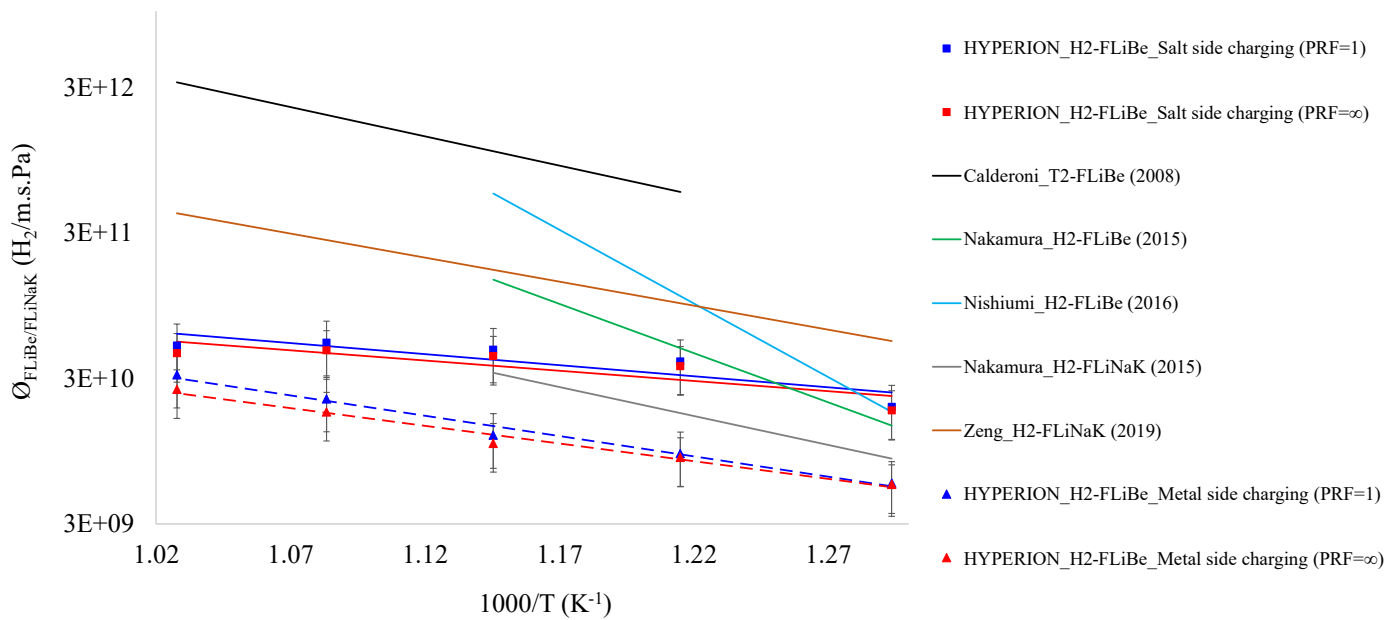

Fig.13(a) Arrhenius plots for $H_2$ permeability in FLiBe estimated from HYPERION experiments over a temperature range of 773 K – 973 K for metal-side charging and salt-side charging, and their comparison with earlier studies

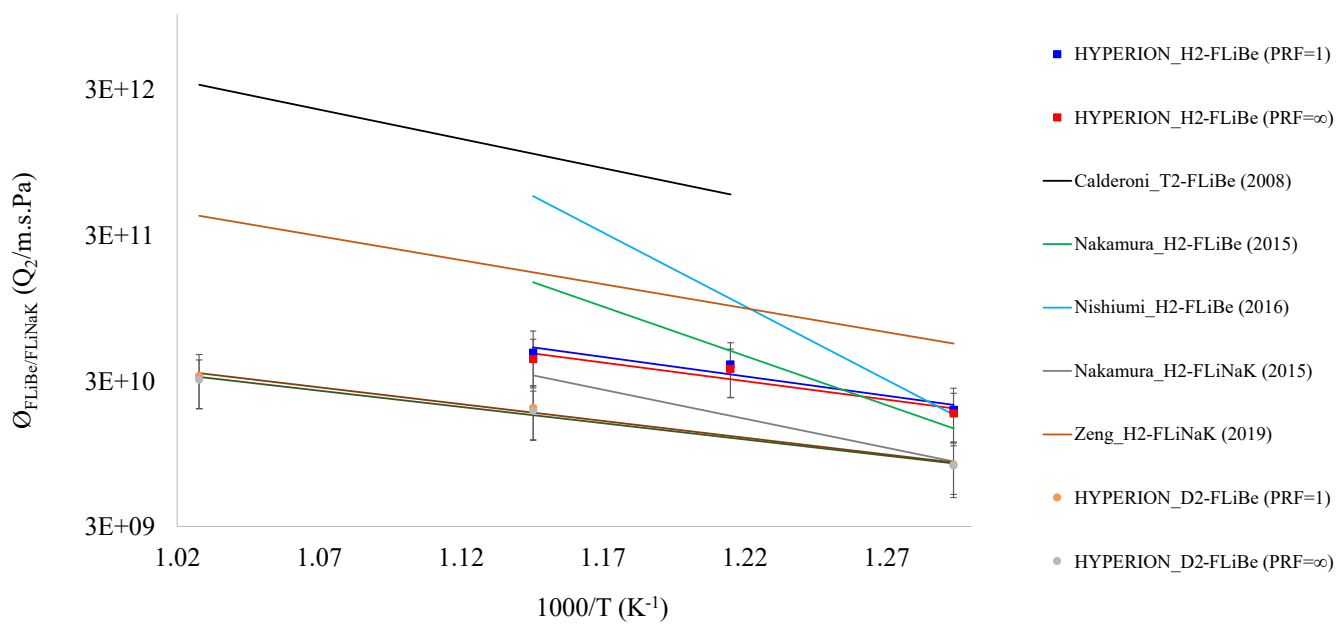

Fig.13(b) Arrhenius plots for $H_2$ and $D_2$ permeabilities in FLiBe estimated from HYPERION experiments over a temperature range of 773 K – 973 K for salt-side charging, and their comparison with earlier studies
(For interpretation of the color-coded data in the legend, please refer to the online version of this article)